\documentclass[a4paper,english,12pt]{article}

\usepackage[version=3]{mhchem} 
\usepackage{fontenc}           
\usepackage{hyperref}
\usepackage[latin1]{inputenc}
\usepackage{float}
\usepackage{color}
\usepackage{graphicx}
\usepackage{amssymb}
\usepackage{babel}
\usepackage{amsmath}
\usepackage{fancyhdr}
\usepackage{geometry}
\usepackage{comment}
\usepackage[normalem]{ulem}
\usepackage{mathtools}   
\usepackage{xcolor}

\usepackage[affil-it]{authblk}

\author[1]{Angus Crookes}
\author[1]{Ben Yuen}
\author[1]{Angela Demetriadou\thanks{a.demetriadou@bham.ac.uk}}
\affil[1]{School of Physics and Astronomy, University of Birmingham, Edgbaston, Birmingham, B15 2TT, United Kingdom}

\title{Taming plasmonic nanocavities for subradiant entanglement}

\begin{document}

\maketitle

\begin{abstract}
Recent rapid advances in quantum nanoplasmonics offer the potential for accessing quantum phenomena at room temperature. Despite this, entangled states have not yet been realised, and remain an outstanding challenge. In this work, we demonstrate how entanglement emerges in plasmonic nanocavities, which are inherently multi-mode, and demonstrate the conditions necessary for entanglement to persist. We find that, in general, these conditions are broken due to coupling with multiple plasmonic modes of different parity. We address this challenge with a new nanocavity design that supports high selective coupling to a single mode, enabling the robust generation of subradiant entanglement in nanoplasmonics. Our results open exciting prospects for leveraging simple plasmonic setups in ambient conditions for applications in quantum communication, sensing and rapid quantum memories. 

. 

\end{abstract}

\thispagestyle{empty}

\section*{Introduction}

Quantum entanglement plays an indispensable role in various fields of quantum information, including quantum computation \cite{nielsen2010quantum}, cryptography \cite{gisin2002quantum,pirandola2020advances}, metrology \cite{giovannetti2011advances} and quantum teleportation \cite{pirandola2015advances}. Inspired by these many applications, quantum devices have seen rapid progression over the past decade, where research has developed a myriad of high fidelity qubits that can be coherently controlled and initialised \cite{laflorencie2016quantum, chatterjee2021semiconductor, wendin2017quantum}. In fact, entanglement across a wide range of platforms has already been achieved~\cite{friis2019entanglement}, such as with trapped ion qubits \cite{turchette1998deterministic, haffner2005scalable}, photon pairs \cite{aspect1982experimental, zukowski1995entangling, wang2016experimental}, through utilizing cold atoms \cite{yang2020cooling}, and between superconducting qubits \cite{riste2013deterministic, steffen2006measurement}. 
However, despite the wide variety of platforms, most have complicated and large setups that operate at cryogenic temperatures to maintain high coherence times and entanglement~\cite{corcoles2011protecting,wang2022towards}.

In contrast, plasmonic nanocavities have enabled the exploration of quantum phenomena at ambient conditions, due to their unprecedented ability to confine light \cite{tame2013quantum, saez2022plexcitonic, gonzalez2011entanglement, moreno2004theory, huck2009demonstration}. Furthermore, their small scale ($\sim80$ nm) has the potential for high density integration into nanoscale devices and the ability to harness chemical processes at the single molecule level~\cite{cortes2020challenges}. Plasmonic nanocavities are also easy to synthesise, and are chemically robust, making experiments reliable and highly reproducible~\cite{simoncelli2016quantitative,chikkaraddy2018mapping, benz2016single, jakob2023giant}. Hence, several studies have demonstrated single molecule strong coupling at room temperature \cite{chikkaraddy2016single, zengin2015realizing, simoncelli2016quantitative,chikkaraddy2018mapping}, which provides a potential avenue for entanglement. Despite this, entangled states have not yet been realised with nanoplasmonics and remains an outstanding challenge.

In this work, we demonstrate how entanglement in realistic, multi-mode plasmonic nanocavities emerges via the plasmonic loss, and obtain the general conditions for its persistence. We show why these conditions are not met in conventional plasmonic systems, since the quantum emitters (QEs) couple to multiple modes that have different parity and relative coupling strengths. 
To address these challenges, we present a unique, easily realizable plasmonic nanocavity that suppresses the influence of multiple modes on entanglement by isolating a single even mode that is strongly coupled to the QEs. 
In this new design, we demonstrate the robust generation of subradiant entanglement via plasmonic loss, which is irrespective of QE placement. Our results open exciting prospects for generating entanglement at room temperature with simple plasmonic setups.

\section*{Results and discussion}
Creating stable quantum entanglement in plasmonic systems is extremely challenging, as unavoidable plasmonic losses such as Ohmic heating and radiation, result in the loss of quantum coherence. However, cavity dissipation can also be used to drive QEs into entangled states~\cite{plenio1999cavity, hughes2005modified, maniscalco2008protecting, francica2009off, hou2014dissipation, tokman2023dissipation, martin2011dissipation,bedingfield2023subradiant}. 
The central idea is to prepare two QEs in a superposition state, where one part evolves through interaction with the plasmonic cavity, while the other part is protected from cavity dissipation and is persistent.

\begin{figure}
	\centering
	\includegraphics[width=1\textwidth]{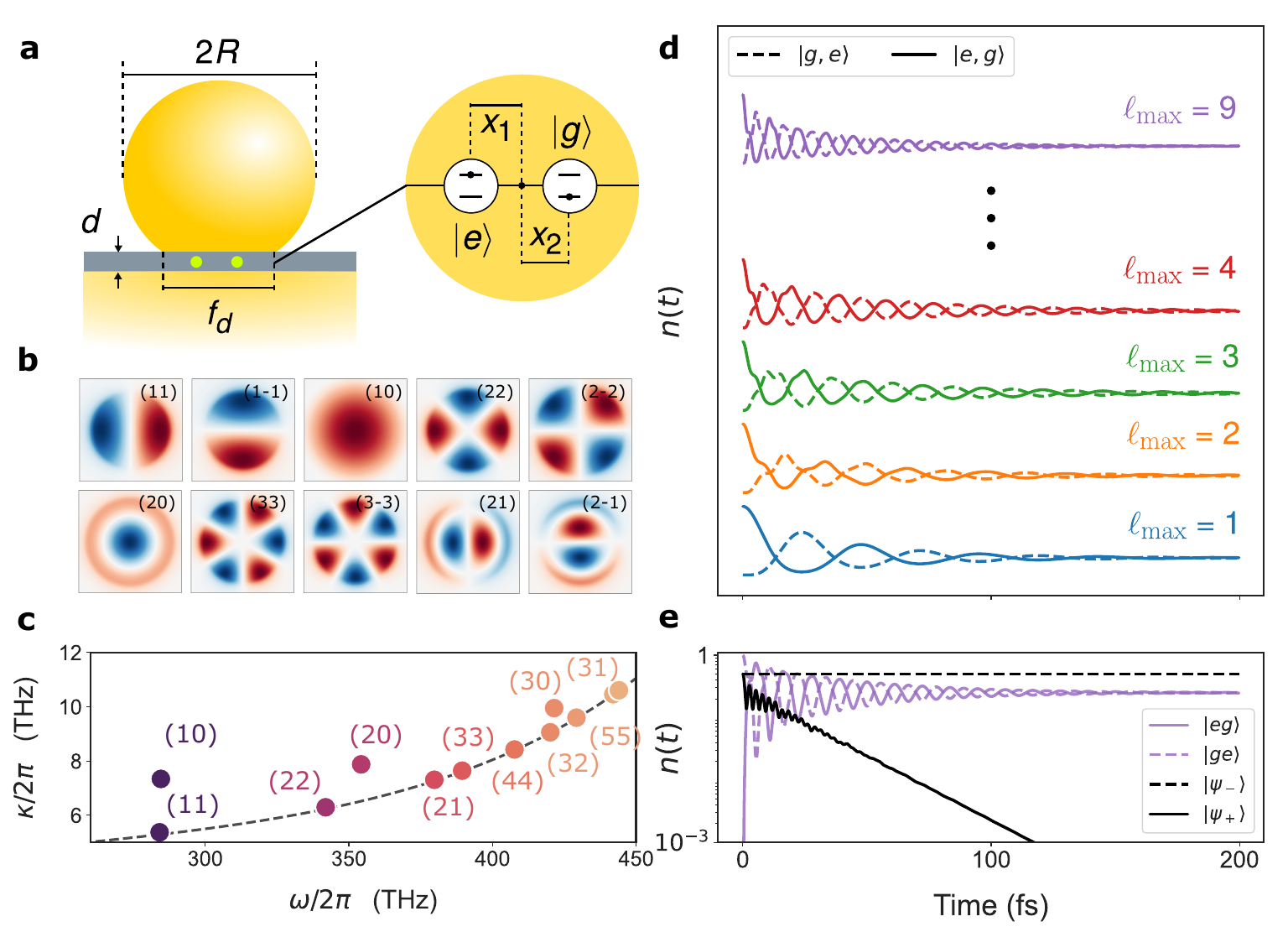}
	\caption{\textbf{Entanglement formation in nanoparticle on mirror cavities.} \textbf{a} Schematic of a gold nanoparticle on mirror (NPoM) cavity with radius $R = 40$ nm, facet diameter $f_{\text{d}}=16$ nm, gap spacing $d_{\text{gap}}=1$ nm, and gap permittivity $n_{\text{gap}}=2.5$.  The cavity is loaded with two quantum emitters at the centre of the nano-gap at position $x_1$ and $x_2$ respectively. \textbf{b} Electric field, $E_{z}(x,y,0)$, of the first fifteen quasinormal modes (QNM's) supported by the bare cavity. \textbf{c} Complex representation of the QNM eigenvalues, $\tilde{\omega}_{\xi} = \omega_{\xi} - \frac{i\kappa_{\xi}}{2}$, with respect to the eigenvalues of a metal-insulator-metal waveguide  (black-dashed line). \textbf{d} Population of each QE's excited state (solid - $|e, g\rangle$  and dashed - $|g, e\rangle $) as a function of time at $x_1 = x_2 = 0$. The interaction with off-resonant plasmonic modes increases the oscillation frequency but despite this always forms the entangled subradiant state, $|\psi_-\rangle=\frac{1}{\sqrt{2}}\left(|e,g\rangle - |g,e\rangle \right)$ \textbf{e} Population of each quantum emitters excited state and of the entangled states $|\psi_-\rangle$ and $|\psi_+\rangle$ for the case of $\ell_{\text{max}}=9$ and $x_1=x_2=0$.
	}
	\label{fig:modes}
\end{figure}

This population trapping mechanism is particularly successful when the QEs are subject to strong field enhancements and large cavity losses, as this increases the rate of entanglement formation. Therefore, recent research has focused on entanglement within nanoparticle on mirror (NPoM) cavities, which are easy to synthesise, very robust, and provide extremely small mode volumes that bring the system into the strong coupling regime at room temperature \cite{chikkaraddy2016single, tokman2023dissipation, bedingfield2023subradiant}. A schematic of the NPoM system is shown in Fig. \ref{fig:modes}a in which both QEs are prepared in the initial state $|e,g\rangle = \frac{1}{\sqrt{2}}\left(|\psi_-\rangle + |\psi_+\rangle\right)$, where $|\psi_{\pm}\rangle=\frac{1}{\sqrt{2}}\left(|e,g\rangle \pm |g,e\rangle \right)$ are both entangled states, yet $|e,g\rangle$ is not. In previous studies~\cite{bedingfield2023subradiant, tokman2023dissipation},
it is assumed that only a single plasmonic mode is interacting with the QEs, which is a common approximation in quantum plasmonics. 
For the single mode case, they demonstrated that when the coupling of the QEs to the plasmonic mode is identical, $|\psi_+\rangle$ exchanges energy with the cavity mode and decays through plasmonic dissipation, while $|\psi_-\rangle$ is decoupled from the cavity (i.e. is subradiant), which remains to form a persistent entangled state. 

However, in reality plasmonic systems feature multiple dissipative and broadband modes that all couple strongly to the QEs in the nanocavity, even if the modes are off-resonant \cite{li2016transformation, medina2021few, he2022principle, yang2024electrochemically}. In this work, we explicitly describe the interaction between multiple plasmonic modes and two QEs using an open quantum system formalism, where the density operator $\rho$ evolves under: 
\begin{equation}
	\dot{\rho}(t) = -i\left[\mathcal{H}_{\text{sys}},\rho\right] + \sum_{\xi}^N\kappa_{\xi}\left(a_{\xi}\rho a_{\xi}^{\dag}-\frac{1}{2}\{a_{\xi}^{\dag}a_{\xi}, \rho\}\right) 
	\label{master_eqn}
\end{equation}
with 
\begin{equation}
	\mathcal{H}= \sum_{\xi=1}^N\omega_{\xi} a_{\xi}^{\dag}a_{\xi} + \sum_{j=1}^M \frac{\omega_e}{2}\sigma_{\text{z}}^{j} +  \sum_{\xi,j}^{N,M} g_{\xi}(\mathbf{r}_j)a_{\xi}^{\dag}\sigma_j + g^*_{\xi}(\mathbf{r}_j)a_{\xi}\sigma_j^{\dag}
	\label{h_sys}
\end{equation}
where $a_{\xi}^{\dag}$ and $a_{\xi}$ are the creation and annihilation operators for each plasmonic mode $\xi$ with frequency $\omega_{\xi}$ and loss rate $\kappa_{\xi}$ and  $\sigma_j^{\dag}$ and $\sigma_j$ are the raising and lowering operators for a QE with transition frequency $\omega_{\text{e}}$. In addition, the position dependent coupling strength between mode $\xi$ and quantum emitter $j$ at $\mathbf{r}_j$ is given by $g_{\xi}(\mathbf{r}_j) = g_{\xi j}e^{i\phi_{\xi j}}$, which is separated into its magnitude, $g_{\xi j}$ and phase, $\phi_{\xi j}$. 

For such multi-mode systems, the population trapping mechanism is satisfied when a dark (subradiant) state $|\psi_{\text{D}}\rangle = \cos\theta|e,g\rangle + e^{i\chi}\sin\theta |g,e\rangle$ is decoupled from the cavity and a bright (super-radiant) state $|\psi_{\text{B}}\rangle = -e^{i\chi}\sin\theta|e,g\rangle + \cos\theta|g,e\rangle$ decays rapidly, where $\theta \in \left[-\pi/2,\pi/2\right]$ and $\chi \in [0,2\pi]$. The state $|\psi_{\text{D}}\rangle$ is dark (and therefore persistent) under the conditions:
\begin{equation}
	\left(\frac{g_{\xi 1}}{g_{\xi 2}}\right)e^{i\left(\phi_{\xi 1}-\phi_{\xi 2}\right)} = -e^{i\chi}\tan\theta ~~~~ \text{for all modes} ~\xi
	\label{h_int_condition}
\end{equation}
which is valid (but not necessarily satisfied) for any dissipative multi-mode cavity, and depends on the magnitude and phase of the QEs' coupling strengths to each mode in the cavity (see Methods for more details). 
Therefore, the initial state $|e,g\rangle = \cos\theta|\psi_{\text{D}}\rangle - e^{i\chi}\sin\theta|\psi_{\text{B}}\rangle$ that is considered here, generates the persistent entangled state $|\psi_{\text{D}}\rangle$ only when Eq. (\ref{h_int_condition}) is satisfied.

To assess the entanglement capabilities of commonly used plasmonic systems using equation~\ref{h_int_condition}, we consider the  NPoM configuration and calculate its multiple modes. We obtain their frequencies, loss rates, and interaction strengths using the auxiliary eigenvalue method \cite{yan2018rigorous, lalanne2018light, qnmeig}. Here, eigenstates of the bare cavity are expressed as quasinormal modes (QNMs) that have a finite lifetime given by their complex eigenfrequency $\tilde{\omega}_{\xi} = \omega_{\xi} - i\frac{\kappa_{\xi}}{2}$, where $\omega_{\xi}$ and $\kappa_{\xi}$ are the resonant frequency and decay-rate corresponding to each mode with electric and magnetic fields $\tilde{\mathbf{E}}_{\xi}(\mathbf{r})$ and $\tilde{\mathbf{H}}_{\xi}(\mathbf{r})$. 
From these fields, one can obtain the coupling strength of each QNM $\xi$ to each QE $j$ as:  $g_{\xi j}e^{i\phi_{\xi j}} = \sqrt{\frac{\omega_{\xi}}{\hbar \text{V}_{\xi} }}\boldsymbol{\mu}\cdot\mathbf{\tilde{E}}_{\xi}(\mathbf{r}_j)$ where $\boldsymbol{\mu}$ is the QE dipole moment and $\text{V}_{\xi}=\iiint_{\Omega} \left[\tilde{\mathbf{E}}\cdot\frac{\partial \omega \boldsymbol{\epsilon}}{\partial \omega}\tilde{\mathbf{E}} - \mu_0 \tilde{\mathbf{H}}\cdot\frac{\partial \omega \boldsymbol{\mu}}{\partial \omega}\tilde{\mathbf{H}}\right]dV$ is the QNM normalisation factor.
The electric field, $\tilde{E}^{\xi}_z(x,y,0)$, of the first ten QNMs at the centre of the nanocavity are shown in Fig. \ref{fig:modes}b., each characterised by an index $\xi=(lm)$, where $l\in \mathbb{Z}^{+}$ is the number of radial anti-nodes and $-l\leq m \leq l$ the pairs of azimuthal anti-nodes \cite{kongsuwan2020plasmonic}. The $m=0$ modes are cylindrically symmetric and have a node at the nanocavity centre, while the $m \neq 0$ modes are degenerate with an orthogonal $m,-m$ pair, and have an anti-node at the nanocavity centre. Importantly, the value of $|m|$ defines the modes parity, with $|m| = 0, 2, 4, 6, ...$ for even and $|m| = 1, 3, 5, 7, ...$ for odd modes respectively~\cite{kongsuwan2020plasmonic, bedingfield2022plasmonic}.
The complex eigenfrequencies associated with each QNM are displayed in Fig.~\ref{fig:modes}c. In this system, all $m\neq 0$ modes predominantly loose energy via ohmic heating, since they closely follow the eigenvalues of a metal-insulator-metal waveguide (i.e. dashed line)~\cite{kongsuwan2020plasmonic} while all $m=0$ modes have an additional radiative loss (see Supplementary Information for details on the calculation). The parameters $\kappa_{\xi}$, $\omega_{\xi}$, and $g_{\xi j}e^{i\phi_{\xi j}}$  are obtained from the mode decomposition for the first 103 modes up to $\xi = (90)$, which allows us to assess the impact of the multiple modes on the entanglement.

The entanglement conditions described by equation (\ref{h_int_condition}) are always satisfied for two QEs at the same position within the nanocavity, as both QEs have identical coupling strengths and phases ($g_{\xi 1}=g_{\xi 2}$ and $\phi_{\xi 1}=\phi_{\xi 2}$) for each mode $\xi$. Therefore, in this case the system always exhibits persistent entanglement irrespective of the number of interacting modes, as the state $|\psi_-\rangle$ is always dark. This is demonstrated numerically in Fig. \ref{fig:modes}d where we consider two QEs at the nanocavity centre i.e. $x_1=x_2=0$, with transition frequency $\omega_e = \omega_{(10)} = 283$ THz and dipole moment $\mu_0= 1\times10^{-28}$ Cm. Here, we calculate the time evolution including interacting modes up to $(\ell_{\text{max}}0)$ for various values of the index $\ell_{\text{max}}$ where only $m=0$ modes contribute,  since  all $m\neq 0$ modes vanish at the centre ($x=y=0$).  
For all values of $\ell_{\text{max}}$, the two QEs remain correlated with one another, and without exception relax to the same subradiant entangled state $|\psi_-\rangle$ with steady state density matrix $\rho_{\text{ss}}=\frac{1}{2}|g,g\rangle\langle g,g| + \frac{1}{2}|\psi_-\rangle\langle\psi_-|$. This is shown for the specific case $\ell_{\text{max}}=9$ in Fig. \ref{fig:modes}e from which it is clear that the population of $|\psi_-\rangle$ does not change, and is therefore decoupled from the cavity. This is consistent with the single mode approximation ($\ell_{\text{max}}=1$), with now more rapid oscillations between the two QEs as we include off-resonant modes.
This highlights that multiple lossy modes do not prevent the formation of entanglement in plasmonic systems, provided the conditions in Eq (\ref{h_int_condition}) are satisfied.

\begin{figure}[h!]
	\centering
	\includegraphics[width=1\textwidth]{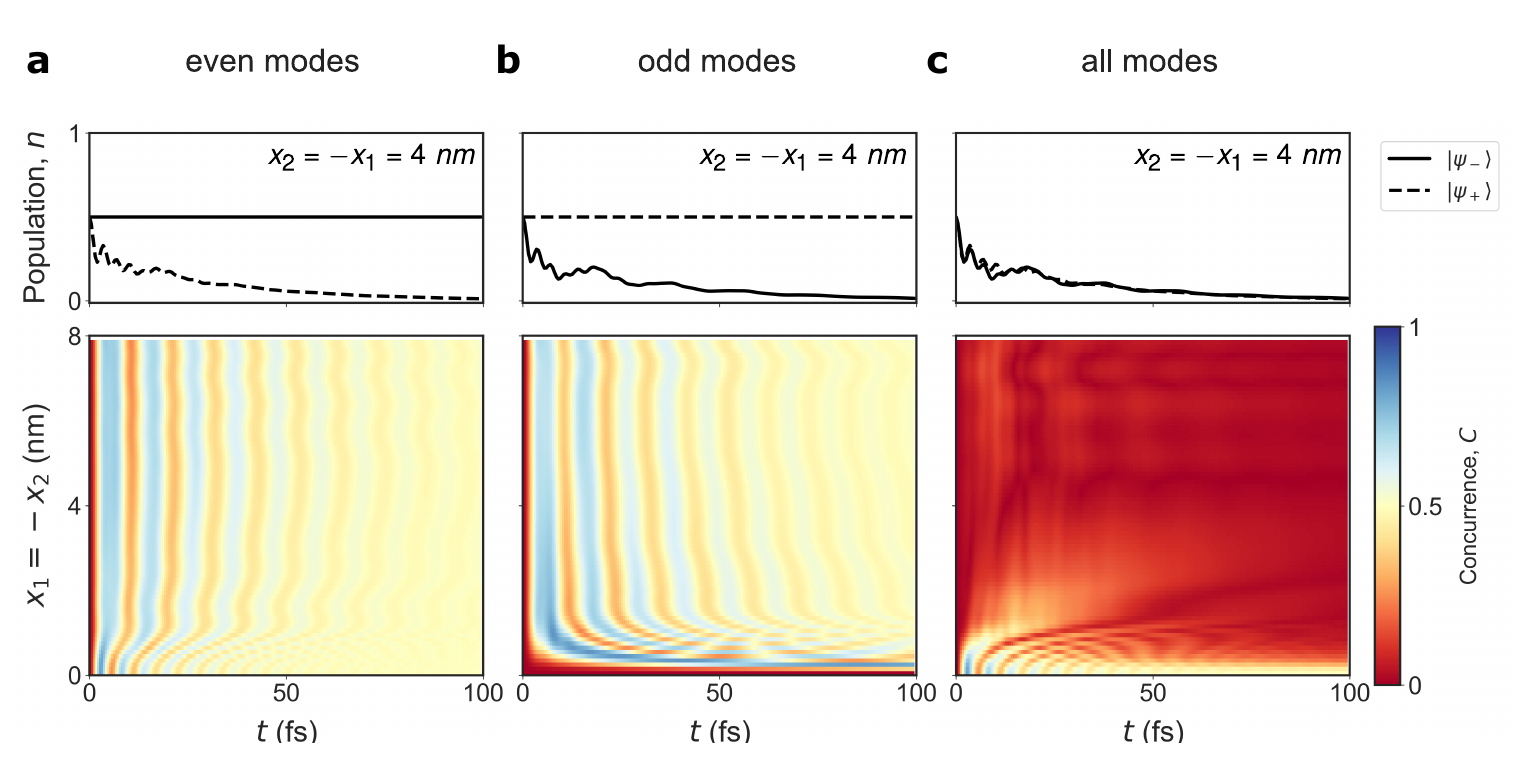}
	\caption{\textbf{Effect of mode parity on sub-radiant entanglement.} \textbf{a-c} Concurrence as a function of symmetric displacement of both quantum emitters (i.e. $x_1 = -x_2$) including only: \textbf{a} even plasmonic modes ($m=0, 2, 4, ...$) \textbf{b} odd plasmonic modes ($m=1,3,5, ...$) and \textbf{c} all modes ($m=0,1,2,3,...$) in the interaction. The corresponding populations of both entangled states $|\psi_-\rangle$ and $|\psi_+\rangle$ are also shown at a displacement of $x_2=-x_1=4$ nm, demonstrating a clear switching between the entangled subradiant states for even and odd modes respectively. No persistent entanglement is observed with both even and odd modes.}
	\label{fig:even_odd_entanglement}
\end{figure}

Whilst this example clearly shows persistent entanglement in a multi-mode plasmonic nanocavity, in reality two QEs cannot be located at the exact same position. 
To guarantee that the magnitude of both QEs' coupling strengths to each mode still satisfies Eq. (\ref{h_int_condition}), we first place them symmetrically away from the centre of the nanocavity (i.e. $x_2=-x_1$ and $g_{\xi 1} = g_{\xi 2}$) such that the phase $\phi_{\xi j}$ of the coupling strength becomes the deciding factor on entanglement. 
In this case, when $x_{1}=-x_{2} \neq 0$, a more expansive set of modes (i.e. $m\neq 0$) that can have different parities now couple to both QEs. Since Eq (\ref{h_int_condition}) shows that persistence of entanglement is dependent on the mode parity ($\phi_{\xi 1}$,~$\phi_{\xi 2}$), we consider three situations where we include only: (1) even modes, (2) odd modes and (3) all modes coupled to the QEs as shown in Fig.~\ref{fig:even_odd_entanglement}.

To better quantify the entanglement for any position of the QEs in the nano-cavity, we use the Wootters concurrence \cite{wootters2001entanglement, walter2016multipartite}:
\begin{equation}
	C(\rho) = \text{max}\left(0, \lambda_1-\lambda_2-\lambda_3-\lambda_4\right)
	\label{concurrence}
\end{equation}
where $\{\lambda_i\}$ is the set of ordered eigenvalues of $R = \sqrt{\sqrt{\rho}\tilde{\rho}\sqrt{\rho}}$ and $\tilde{\rho} = \sigma_y\otimes\sigma_y\rho^*\sigma_y\otimes\sigma_y$ is the spin flipped reduced density matrix. The entanglement between both QEs is then determined by performing a partial trace over the plasmonic degrees of freedom and calculating the concurrence on the remaining reduced density matrix. In this way, a maximally entangled state is given by a concurrence of unity while a completely separable state causes the concurrence to vanish. Fig. \ref{fig:even_odd_entanglement} a. shows the entanglement when only even plasmonic modes ($m=0, 2, 4, ...$) are coupled with the QEs; the resulting steady state remains entangled due to a population of the state $|\psi_-\rangle$. In Fig. \ref{fig:even_odd_entanglement}b only odd modes ($m=1,3,5, ...$) are coupled to the QEs, which again results in a steady state that exhibits entanglement but now due to a population of the state $|\psi_+\rangle$. 
In this system, when $x_1=-x_2$ and therefore $g_{\xi 1} = g_{\xi 2}$, the conditions for persistent entanglement due to population of the dark state $|\psi_{D}\rangle$ described in Eq.~(\ref{h_int_condition}) can be reduced to:
\begin{equation}
	e^{i\chi} = -e^{i\left(\phi_{\xi 1}-\phi_{\xi 2}\right)},
	\label{entanglement_conditions}
\end{equation}
with $\theta = \pi/4$, which must be satisfied for all modes independently for entanglement to emerge.
It is clear from Eq. (\ref{entanglement_conditions}) that the solutions (1) $\chi = \pi$ is satisfied when all the nanocavity modes are even since $\phi_{\xi 1}-\phi_{\xi 2}=0$ and (2) $\chi=0$ is satisfied when all the modes in the nanocavity are odd, since $\phi_{\xi 1}-\phi_{\xi 2} = \pi$.
In each case, the corresponding entangled state protected from dissipation is switched from $|\psi_-\rangle$ for even modes to $|\psi_+\rangle$ for odd modes, as seen in the top panels of Fig \ref{fig:even_odd_entanglement}a-b. 

\noindent If both QEs are strongly coupled to a combination of even and odd modes, then the persistence conditions in Eq. (\ref{entanglement_conditions}) are not satisfied, and no entanglement is observed at long times. Fig. \ref{fig:even_odd_entanglement}c shows exactly this - when the complete collection of plasmonic modes is included (both even and odd modes) there is a complete loss of entanglement, with $C$ vanishing for all $x_1 = -x_2 > 0.5$ nm.  This is emphasised in the top panel, showing both populations of $|\psi_{-}\rangle$ and $|\psi_{+}\rangle$ decay at $x_2=-x_{1}=4$ nm. 
In addition, we so far considered symmetric QE placement to ensure that $g_{\xi 1} = g_{\xi 2}$. However, if the ratio $g_{\xi 1}/g_{\xi 2}$ is different for each mode $\xi$, which is the case for asymmetric QE placement, then entanglement is also completely removed. This is because the persistence conditions in Eq. (\ref{h_int_condition}) are not satisfied for asymmetric QE placement in most plasmonic nanocavities (see Supplementary Information).
Hence, both the relative coupling strengths of the two QEs to each mode, as well as the mode parity, dominate the formation of subradiant entangled states in multi-mode cavities. This presents a major challenge in the generation of entanglement within plasmonic nanocavities, where modes of different parity and strong field enhancements, almost always co-exist.

\subsection*{Taming plasmonic systems for entanglement}

\begin{figure}[h!]
	\centering
	\includegraphics[width=1\textwidth]{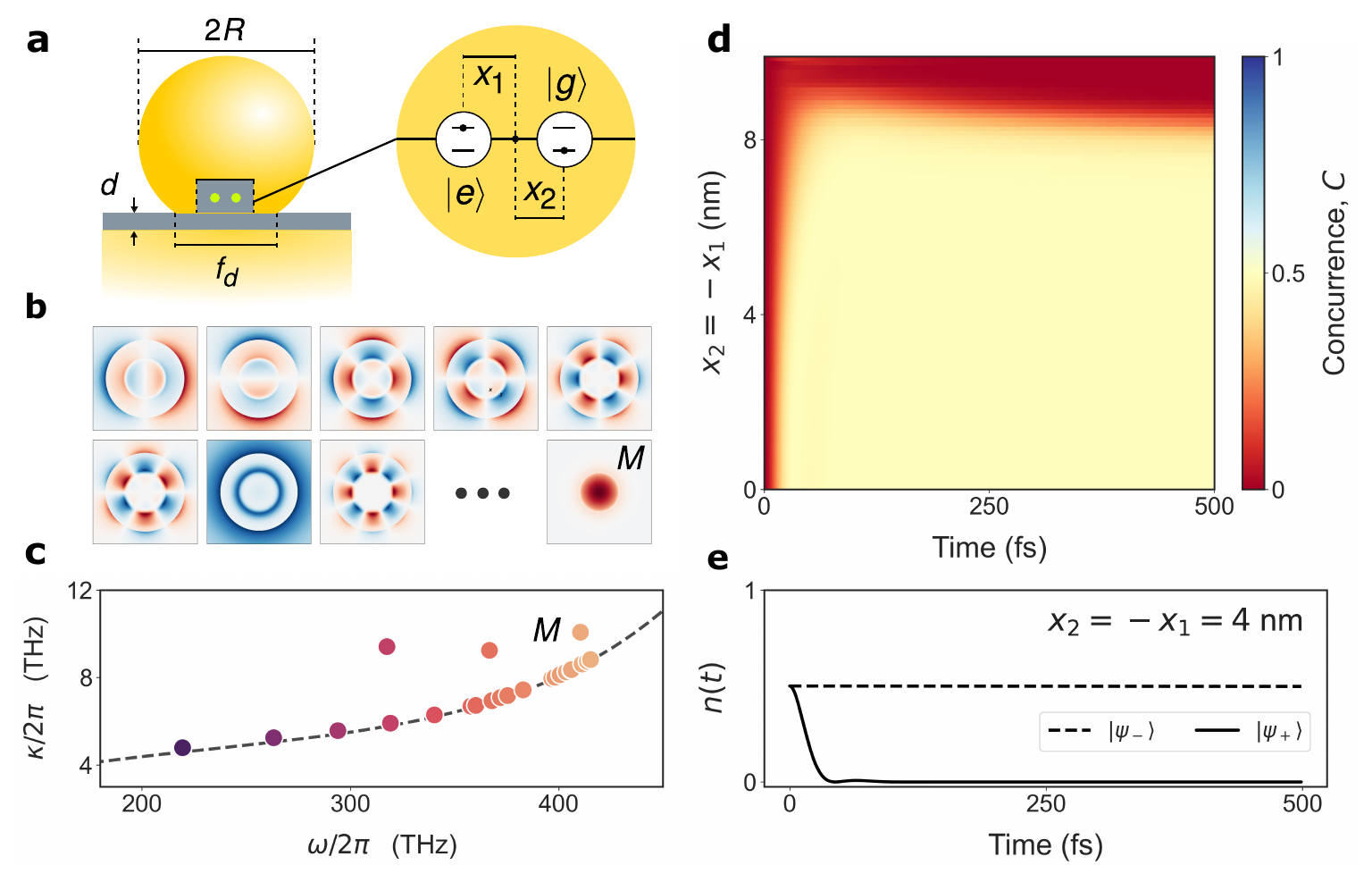}
	\caption{\textbf{Plasmonic nanocavity design for robust entanglement} \textbf{a} Schematic of a gold nanoparticle on mirror (NPoM) cavity with radius $R = 40$ nm, facet diameter $f_{\text{d}}=30$ nm, gap spacing $d_{\text{gap}}=1$ nm, and gap permittivity $n_{\text{gap}}=2.5$. Into the nanoparticle facet a hollow cylinder is etched, of height $h = 3$ nm and radius $r = 10$ nm. \textbf{b} Electric field distribution of quasinormal modes (QNMs) within the perforation at $z=(d_{gap}+h)/2$. \textbf{c} Complex representation of the QNM eigenvalues with respect to those of a metal-insulator-metal waveguide (black-dashed line) \textbf{d} Concurrence as a function of time and displacement ($x_2 = -x_1$) of both emitters. \textbf{e} Quantum emitter entangled state populations, $|\psi_-\rangle$ and $|\psi_+\rangle$, highlighted at a displacement of $x_2=-x_1=4$ nm. }
	\label{fig:new_design_entanglement}
\end{figure}

To overcome this problem, one needs to design plasmonic nanocavities where the system favours either even or odd  modes in a certain spatial region to place the QEs. 
To achieve this, we introduce a new design that fulfils this parity selection. A schematic of our new design is shown in Fig. \ref{fig:new_design_entanglement}a, where we consider the same nanoparticle-on-mirror, but now with a hollow cylinder of radius $r = 10$ nm and height $h=3$ nm etched into the nanoparticle facet of size $f_{\text{d}}=30$ nm. The $z$-component of the electric field of each QNM, $\tilde{E}^{\xi}_z(x,y,z=(d_{\text{gap}}+h)/2)$, is shown inside the hollow cylinder in Fig. \ref{fig:new_design_entanglement}b.
This unique nanocavity design also supports both even and odd modes, but the fields of $m\neq 0$ modes are now concentrated at the edges of the hollow cylinder and have very small coupling strengths inside. Instead, the fields within the cylinder are dominated by a single (even) mode at $\lambda = 730$ nm (labelled $M$). This mode originates from the cylindrical hollow geometry and has a field enhancement over an order of magnitude larger than all other modes within this spatial region. The complex eigenfrequencies associated with each QNM are displayed in Fig. \ref{fig:new_design_entanglement}c and again share close agreement with those obtained from a metal-insulator-metal waveguide (dashed line). Importantly, despite overlapping resonances in the frequency domain, we are able to achieve highly selective coupling of QEs to only mode $M$ by placing them within the hollow cylinder, where the fields of all other modes are suppressed.

The entanglement as a function of time and symmetric placements ($x_2=-x_1$) at $z=(d_{\text{gap}}+h)/2$ is shown in Fig. \ref{fig:new_design_entanglement}d-e. where both QEs have dipole moment $\mu_0$, frequency $\omega_{e} = \omega_{M} = 730 $ nm and each interacts with $40$ plasmonic modes. 
In this new design, although the persistence conditions in Eq. (\ref{h_int_condition}) are not strictly satisfied, the weak coupling of all other modes that violate these conditions are negligible, and thus have little effect on the dynamics over which entanglement emerges. Therefore, this allows for persistent entanglement to emerge for $x_2=-x_1\lesssim 8$ nm as a result of the persistence of $|\psi_{-}\rangle$.
Beyond $8$nm from the centre, other modes become significant due to the high charge confinement on the hollow cylinder walls, which break the entanglement conditions in this region (Fig.~\ref{fig:new_design_entanglement}d). However, this is just $1-2$ nm away from the metallic wall of the hollow cylinder.

\begin{figure}[h!]
	\centering
	\includegraphics[width=0.7\textwidth]{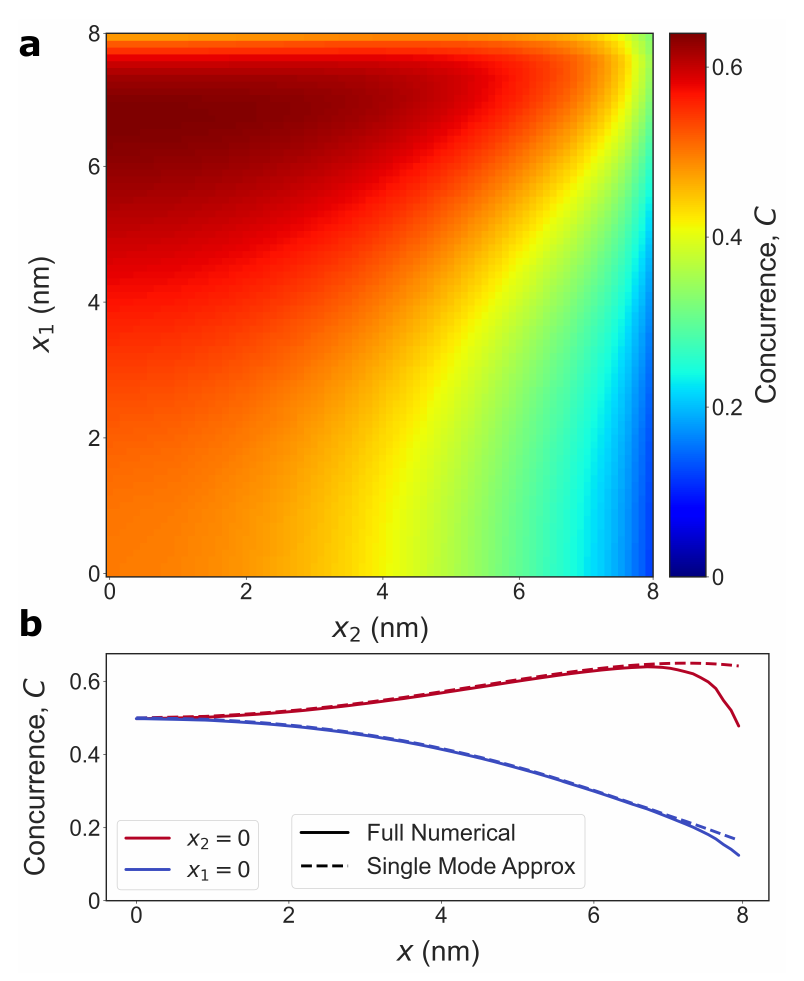}
	\caption{\textbf{Entanglement generation under asymmetric displacement.} \textbf{a} Concurrence, evaluated at $t=1$ ps, as a function of asymmetric quantum emitter position, $x_1$ and $x_2$. The QE at $x_1$ is initially excited. \textbf{b} Concurrence, evaluated at $t=1$ ps, as a function of quantum emitter position for $x_1=0$ (blue) and $x_2=0$ (red). Calculations are performed using forty modes (solid lines) and a single mode approximation (dashed lines). }
	\label{fig:asymmetric_entanglement}
\end{figure}

Due to the unique performance of our new nanocavity design, entanglement is actually also persistent even for asymmetric QE positions. 
Fig. \ref{fig:asymmetric_entanglement}a shows the concurrence after a long time has elapsed (i.e. $t=1$ ps) as a function of both QE positions ($x_1$ and $x_2$) considering coupling to 40 plasmonic modes. In contrast to other plasmonic cavities, entanglement is observed at all positions within the hollow cylinder for $x_1, x_2 \lesssim 8$ nm.
Entanglement between QEs with asymmetric positions is rarely possible in other nanocavities, because the multi-mode persistence conditions in Eq. (\ref{h_int_condition}) are not satisfied when $g_{\xi 1}/g_{\xi 2}$ is different for each mode $\xi$. Instead, the persistence of entanglement here originates from the strong coupling of both QEs to a single plasmonic mode, which allows us to safely take a single-mode approximation, within which the steady state density matrix can be expressed as:
\begin{equation}
	\rho_{\text{ss}} = \left(1-\frac{1}{1+|\alpha|^2}\right)|g,g\rangle\langle g,g| + \frac{1}{1+|\alpha|^2}|\psi_{\text{D}}\rangle\langle \psi_{\text{D}}|
	\label{final_vec}
\end{equation}
where $|\psi_{\text{D}}\rangle = \frac{1}{\sqrt{\tilde{1+|\alpha|^2}}}\left(|e,g\rangle - \alpha|g,e\rangle \right)$ is the (dark) entangled state between both emitters and $\alpha= g_{11}e^{i\phi_{11}}/g_{12}e^{i\phi_{12}}$ is the ratio of coupling strengths between each QE and the spatially isolated plasmonic mode $M$ (see Methods for full details). In this case, there is always a persistent entangled state, which depends on the coupling strengths of both QEs to the plasmonic mode.  Note that when the coupling is symmetric (i.e. $\alpha = 1$) Eq (\ref{final_vec}) reduces to the previous case, where entanglement is due to the state $|\psi_{-}\rangle$. 

The single mode approximation provides an excellent description of entanglement within this plasmonic nanocavity for QEs placed within the hollow cylinder. This can be seen in Fig. \ref{fig:asymmetric_entanglement}b where the concurrence at $t=1$ ps is plotted using the single-mode approximation (dashed lines) together with the full calculations (solid lines) for $x_1=0$ and $x_2=0$ respectively. Furthermore, Eq.~\ref{final_vec} also gives a simple explanation for differences in the degree of entanglement which results from the asymmetric initial state $|e,g\rangle$ and two competing forces: (1) population of the entangled state $|\psi_{\text{D}}\rangle$ i.e. $\langle \psi_{\text{D}}| \rho_{\text{ss}} |\psi_{\text{D}}\rangle = \frac{1}{1+|\alpha|^2}$ and (2) closeness (fidelity) of the entangled state $|\psi_{\text{D}}\rangle$ to the maximally entangled state $|\psi_-\rangle$ i.e. $F = |\langle \psi_-|\psi_{\text{D}}\rangle|^2 = \frac{|1+\alpha|^2}{2(1+|\alpha|^2)}$. When $|\alpha|>1$ (which corresponds to displacement of the initially unexcited QE away from the centre) the population and fidelity of the entangled state $|\psi_{\text{D}}\rangle$ decreases, leading to a decrease in entanglement. However, when $|\alpha|<1$ (which corresponds to displacement of the initially excited QE away from the centre) the fidelity of the entangled state also decreases but the population of $|\psi_{\text{D}}\rangle$ increases more rapidly. This produces an initial increase in the overall entanglement of the system before eventually decreasing at larger separations (see Supplementary Information for more details). 

\section*{Conclusions}

In conclusion, our work elucidates the crucial role of mode parity and relative coupling strength on the generation and manipulation of quantum entanglement of two QEs in cavities. We have shown that the presence of multiple dissipative modes does not prevent the persistence of entanglement. Instead, for cavities with multiple modes, such plasmonic nanocavities, their parity and relative coupling strengths defines the persistence or destruction of entangled states. To address this challenge, we design a new plasmonic nanocavity that suppresses coupling of QEs to all but a single dominant mode, enabling the emergence of robust entanglement regardless of the QE placement. Our results pave the way towards the experimental realization of entanglement in plasmonic systems at ambient conditions, and marks a significant step towards practical applications in quantum communications, sensing and rapid quantum memories using nanoplasmonics.


\newpage

\section*{Methods}

\subsection*{Multi-mode persistence conditions} 

In a multi-mode cavity, the state $|\psi_{\text{D}}\rangle = \cos\theta|e,g\rangle + e^{i\chi}\sin\theta |g,e\rangle$ is subradiant if it is decoupled from the cavity. This occurs when it is part of the null-space of the Hamiltonian i.e. $\mathcal{H}|\psi_{\text{D}}\rangle=0$ which can be expressed as:

\begin{equation}
	\mathcal{H}|\psi_{\text{D}} \rangle = \sum_{\xi=1}^N \left[g_{\xi 1}e^{i\phi_{\xi 1}}\cos\theta + g_{\xi 2}e^{i\phi_{\xi 2}}e^{i\chi}\sin\theta\right] a_{\xi}^{\dag} |0\rangle^{\otimes N}|g,g\rangle = \bar{0}
	\label{null_space_condition}
\end{equation}
where $\bar{0}$ is the null-vector. Importantly, since each term in the sum is independent, the expression in square brackets must vanish for every mode $\xi$ that is coupled to the QEs. Applying this requirement gives Eq. (\ref{h_int_condition}).

\subsection*{Single-mode approximation}

Let an arbitrary state between two quantum emitters be described by the state vector $|\psi_{\text{D}}\rangle = a|e,g\rangle + b|g,e\rangle $. If we require $\mathcal{H}|\psi_{\text{D}}\rangle=0$ then 

\begin{equation}
	\mathcal{H}|\psi_{\text{D}} \rangle = \sum_{\xi=1}^N \left[a g_{\xi 1}e^{i\phi_{\xi 1}} + b g_{\xi 2}e^{i\phi_{\xi 2}}\right] a_{\xi}^{\dag} |0\rangle^{\otimes N}|g,g\rangle = 0
	\label{h_int_app}
\end{equation}
\vspace{0.5mm}

\noindent where $g_{\xi,j}$ is the magnitude of the coupling strength between mode $\xi$ and atom $j$ and $\phi_{\xi j}$ the corresponding phase. In particular, under a single mode approximation we can derive a state that is always part of the null-space of $\mathcal{H}$ such that $|\psi_{\text{D}}\rangle\in \{|\psi\rangle ~|~ \mathcal{H}|\psi\rangle=0 \}$ and

\begin{equation}
	|\psi_{\text{D}}\rangle = \frac{1}{\sqrt{\tilde{N}}}\left(|e,g\rangle - \alpha|g,e\rangle \right)
	\label{null_space_vec}
\end{equation}
\vspace{0.5mm}

\noindent where $\alpha = \frac{g_{11}e^{i\phi_{11}}}{g_{12}e^{i\phi_{12}}}$ and $\tilde{N} = 1 + |\alpha|^2$. Orthonormal to $|\psi_{\text{D}}\rangle$ is the state $|\psi_{\text{B}}\rangle = \frac{1}{\sqrt{1+1/|\alpha|^2}}\left(|e,g\rangle + \frac{1}{\alpha^*}|g,e\rangle \right)$ where $\mathcal{H}|\psi_{\text{B}}\rangle \neq 0$. If the system is initialised in an asymmetric state $|e,g\rangle$ then by transforming to the basis  $\left\{|\psi_{\text{D}}\rangle, |\psi_{\text{B}}\rangle\right\} $ we get:

\begin{equation}
|e,g\rangle = \frac{1}{\sqrt{1+|\alpha|^2}}|\psi_{\text{s}}\rangle + \frac{1}{\sqrt{1+\frac{1}{|\alpha|^2}}}|\psi_{\text{d}}\rangle 
\label{initial_vector}
\end{equation}
\vspace{0.5mm}

\noindent of which part, the $|\psi_{\text{D}}\rangle$ component, is protected from dissipation, while the $|\psi_{\text{B}}\rangle$  component eventually decays to the ground state through plasmonic loss. Therefore, the general steady state density matrix of two quantum emitters interacting with a single mode plasmonic cavity can be written as:

\begin{equation}
	\rho_{\text{ss}} = \left(1-\frac{1}{1+|\alpha|^2}\right)|g,g\rangle\langle g,g| + \frac{1}{1+|\alpha|^2}|\psi_{\text{D}}\rangle\langle \psi_{\text{D}}|
	\label{initial_rho}
\end{equation}
\vspace{0.5mm}

\noindent where $\alpha$ encodes the coupling of each quantum emitter to a single plasmonic mode at any position within the nano-cavity (for more details see Supplementary Information).


\section*{Acknowledgements (not compulsory)}

AD gratefully acknowledges support from the Royal Society University Research Fellowship URF\textbackslash R1\textbackslash 180097 and URF\textbackslash R\textbackslash 231024, Royal Society Research grants RGS \textbackslash R1\textbackslash 211093, funding from ESPRC grants  EP/Y008774/1 and EP/X012689/1.
AD, BY acknowledge support from Royal Society Research Fellows Enhancement Award RGF \textbackslash EA\textbackslash 181038, and AD, AC acknowledge funding from EPSRC for the CDT in Topological Design EP/S02297X/1.

\section*{Author contributions statement}
A.C. conducted the calculations, analysed the results, put the together the numerical codes and performed the analytical derivations. B.Y.assisted with the theoretical methodology and code analysis and A.D. conceived the theoretical investigation on this topic.
All authors reviewed the manuscript. 

\section*{Additional information}

The authors declare no competing interests.

\end{document}